\documentclass[11pt]{article}

\usepackage[letterpaper, margin=1in]{geometry}
\usepackage[T1]{fontenc}
\usepackage[utf8]{inputenc}
\usepackage{newpxtext,newpxmath}
\usepackage[scaled=0.92]{inconsolata}
\usepackage{microtype}
\usepackage{setspace}
\usepackage{amsmath}  %
\usepackage{booktabs}
\usepackage{multirow}
\usepackage{graphicx}
\usepackage{xcolor}
\usepackage{tikz}
\usetikzlibrary{arrows.meta,positioning,fit,backgrounds,calc}
\usepackage{pgfplots}
\pgfplotsset{compat=1.17}
\usepackage{enumitem}
\setlist{itemsep=2pt, topsep=4pt}
\usepackage{caption}
\DeclareCaptionLabelSeparator{bar}{ $|$ }
\usepackage[round,authoryear]{natbib}
\usepackage{fancyhdr}
\usepackage[hidelinks]{hyperref}
\definecolor{linkblue}{RGB}{0,0,190}
\hypersetup{colorlinks=true, linkcolor=linkblue, citecolor=linkblue, urlcolor=linkblue,
  pdftitle={GRP v0.1 Technical Report}, pdfauthor={GRP Team}}

\fancypagestyle{titlepage}{%
  \fancyfoot[L]{\parbox[t]{\dimexpr\textwidth-2em\relax}{\footnotesize\normalfont
    The designation v0.1 reflects our assessment of the progress made so far toward
    end-to-end generative recommendation and the substantial work that remains.}}%
}

\newcommand{\grp}{GRP}
\newcommand{\mgrpo}{mGRPO}

\begin{document}

\thispagestyle{titlepage}
\begin{center}
  {\LARGE\bfseries GRP v0.1 Technical Report\par}
  \vspace{1.4em}
  {\large\bfseries GRP Team\par}
  {\normalsize Snap Inc.\par}
  \vspace{0.3em}
  {\small Contributors are listed in Appendix~\ref{app:contrib}.\par}
\end{center}
\vspace{0.8em}

\noindent
End-to-end (E2E) generative recommendation replaces the multi-stage cascade of
retrieval, ranking, and re-ranking models that powers today's industrial
recommenders with a single model that \emph{generates} the next item's
identifier. Early industrial systems such as OneRec have shown that the paradigm
is viable, yet in our experiments, attempts to replace a production cascade
\emph{overnight} produced negative online results, and for structural reasons. A mature
cascade embeds thousands of granular optimizations accumulated over a
decade---per-source freshness rules, eligibility filters, calibration layers,
diversity controls---that a bare generator does not reproduce; generative
retrieval alone rarely matches the cascade's ranking capability; and on a fast-churning catalog a frozen
generator's exact-item recall collapses within days. A head-to-head swap
therefore measures the absence of those optimizations more than the quality of
the generator, and it offers no signal about which gap to close first.

We instead propose a \emph{progressive} path. First, deploy the E2E model as one
retrieval source among the existing ones and benchmark it against every
incumbent source on the same recommendation metrics for the same users---any shortfall
pinpoints the bottleneck that blocks replacement. Second, retire the sources it
beats, reassign their quota to the model, and grow that quota as the model
improves. Third, let a growing portion of its candidates bypass the early-stage
and eventually the late-stage rankers on the model's own ordering, until the
rankers no longer change the served slate. Every step is independently
measurable and shippable, and the comparison is always made on the incumbent
system's own terms.

This path places requirements on the model itself: it must rank as well as it
retrieves, it must be steerable toward business value, and it must serve inside a
live feed's latency budget. \textbf{\grp{}}, Snap's \textbf{G}enerative
\textbf{R}ecommendation \textbf{P}aradigm, is a single encoder--decoder model
built to meet them: it (i) generates multimodal Semantic IDs as its
\emph{retrieval} output, (ii) scores generated candidates with a jointly-trained
multi-head prediction (MHP) module as its \emph{ranking} output, and (iii) reuses
that same frozen MHP module as the \emph{reward model} for reinforcement-learning
post-training. The report describes the methodology at each stage:
\begin{enumerate}[leftmargin=1.6em, itemsep=1pt]
\item \textbf{Architecture and pre-training.} The strategy is to unify retrieval
  and ranking in one model without letting the two objectives interfere: an
  encoder--decoder trunk decodes a slate of Semantic IDs block-wise and
  independently, while a ranking module is trained jointly but through a
  stop-gradient path, so it can score any candidate without touching the
  generative representation. The pre-training recipe addresses history length, computational cost, and event
  selection by compressing each history item to one token so long histories fit,
  moving dense capacity from the long-input encoder to the decoder, and reserving
  history slots for sparse explicit actions that a recency cap would discard.
\item \textbf{Reward-guided post-training.} The strategy is to reuse the model's
  own frozen ranking module as the reward and to treat reinforcement learning as
  steering that must not erode what pre-training learned. This requires a reward
  that actually distinguishes a single user's candidates, which we obtain by
  redesigning how candidates enter the ranking module, and an objective that
  protects recall: \textbf{\mgrpo{}} adds to GRPO a one-sided, reference-anchored
  margin that activates only when the policy begins to trade a logged target away
  for a higher-reward sample.
\item \textbf{Serving.} The strategy is to keep the served model minimal and to
  slot it behind the existing funnel so it can be introduced progressively. The inference path combines the encoder, block-wise decoder, and ranking module;
  catalog freshness is handled off the hot path by an asynchronously refreshed
  Semantic-ID-to-item catalog; and a configurable bypass of the existing ranking
  stages controls how far the generative candidates travel on their own scores.
  KV-cached, graph-captured decoding brings the path within a live feed's latency
  budget.
\end{enumerate}
Online, as one retrieval source among many, the post-trained \grp{}
configuration with early-ranking bypass delivers strong completion and sharing
alongside above-average watch time.
Reinforcement learning can optimize selected objectives: with view
time as the sole reward, RL improves view time by $+0.45\%$ over the
SFT model, and a larger per-request decode budget yields a
$+0.39\%$ view-time gain with no ranking bypass. Replacing
lower-performing retrieval
sources with \grp{} yields $+0.82\%$ view time and $+2.56\%$ shares on the
short-video surface with neutral platform-level guardrails. We close with the
limitations that separate today's system from the fully end-to-end goal.

\newpage
\tableofcontents
\newpage

\section{Introduction}
\label{sec:intro}

Generative AI has reshaped recommender-system research by making
\emph{generation} rather than \emph{scoring} the primitive: instead of maintaining
a retrieval index followed by a cascade of ranking and value models---often four
or more stages, each its own codebase, training pipeline, and feature set---a
single sequence model consumes a user's interaction history and generates the
identifier of the next item directly. TIGER~\citep{tiger2023} introduced Semantic
IDs and generative retrieval; HSTU~\citep{hstu2024} demonstrated
trillion-parameter scaling of generative transducers; OneRec and
OneRec-V2~\citep{onerec2024,onerec_tr2025,onerecv2_2025} showed a
production encoder--decoder generative recommender aligned with preference
signals; and UniPinRec~\citep{unipinrec2026}, GPR~\citep{gpr2025},
PLUM~\citep{plum2026}, and Gryphon~\citep{gryphon2026} have since pushed the
paradigm toward unified retrieval-plus-ranking at industrial scale. Appendix~\ref{app:related}
positions \grp{} against each of these in detail.

Despite this progress, fully end-to-end (E2E) generative recommendation remains
a research objective with few successful replacements of production cascades. In
our own experiments, direct head-to-head replacements produced negative online
results, and our reading is that they share a cause: the attempt to replace a
multi-stage cascade with a single model \emph{overnight}. A mature cascade embeds thousands of granular
optimizations---per-source freshness rules, eligibility filters, calibration
layers, diversity controls---that a bare generator does not reproduce, and
comparing the two head-to-head measures the absence of those optimizations more
than the quality of the generator. Three obstacles recur in our own experience:

\begin{enumerate}[leftmargin=1.8em]
\item \textbf{Ranking remains critical.} Generative retrieval produces coarse
  ``next-item'' predictions; a single generated Semantic ID can map to multiple concrete
  items, and the raw generative ordering is far from a good final ordering.
  Applying a downstream ranker on top of items retrieved from the generator
  materially improves view time in our system.
\item \textbf{The catalog is non-stationary.} On a fast-churning short-video
  surface the watch catalog turns over almost completely day-over-day, so a frozen
  model's exact-item hit rate collapses within days as fresh items fall out of
  vocabulary (Section~\ref{sec:freshness}).
\item \textbf{Alignment needs a reward that generalizes to generated candidates.}
  Steering the generator toward business value rather than imitation requires a
  reward signal, and a reward model that ranks \emph{generated} (not just logged)
  candidates is itself a hard modeling problem (Section~\ref{sec:reward-fix}).
\end{enumerate}

\paragraph{Our approach.}
\grp{} addresses these obstacles with one modeling decision and one deployment
decision. The modeling decision is to fold ranking and reward modeling
\emph{into} the generative model: a jointly-trained multi-head prediction (MHP)
module scores generated candidates through a detached, serving-safe input path,
and the same frozen module is reused as the reward for reinforcement-learning
(RL) post-training (Sections~\ref{sec:pretrain}--\ref{sec:posttrain}). The
deployment decision is to be \emph{progressive} (Section~\ref{sec:progressive}):
\grp{} first enters production as an additional retrieval source, is
benchmarked against every incumbent source on the same recommendation metrics, and
receives more traffic where the evaluation supports expansion---replacing
weaker sources, then growing its quota, and eventually bypassing downstream
rankers. This approach evaluates E2E generative recommendation through a sequence of
measurable, individually deployable changes and identifies the remaining gaps to
full replacement.

\paragraph{Contributions.}
This report documents the technical work behind that path:
\begin{enumerate}[leftmargin=1.8em]
\item \textbf{A unified generative recommender and its pre-training recipe}
  (Section~\ref{sec:pretrain}). \grp{} couples an encoder--decoder trunk that
  generates Qwen3-VL Semantic IDs with block-wise-independent target decoding,
  variable-length attention over long user histories, and a jointly-trained MHP
  ranking module whose inputs are detached from the trunk. We show that item-level history fusion accommodates several times more history
  at similar computational cost, that rebalancing dense layers from the encoder
  to the decoder is a Pareto win, improving both accuracy and throughput, and that
  giving rare explicit actions their own reserved
  slice of the history---rather than letting abundant passive watches crowd them
  out---materially lifts reward-weighted recall.
\item \textbf{Reward-guided post-training} (Section~\ref{sec:posttrain}).
  We show that the MHP reward of a naively trained model is nearly
  candidate-indifferent, trace this to a cross-attention layer that had learned to
  ignore the candidate, and redesign the module so candidates enter it through a
  residual path, which makes the heads substantially more discriminative. We
  then propose \textbf{\mgrpo{}}, which adds a one-sided, reference-anchored
  margin to GRPO so that reward optimization cannot erode the base model's recall
  of logged targets: reward-weighted recall improves with recall held, where
  vanilla GRPO is flat on reward and loses recall.
\item \textbf{A production serving path} (Section~\ref{sec:deploy}). The inference path runs the encoder, block-wise decoder, and MHP scorer per request,
  backed by an asynchronously refreshed Semantic-ID$\rightarrow$item catalog and
  configurable ranking-stage bypass; KV caching, CUDA-graph capture, C++
  preprocessing, and a batched reverse lookup cut end-to-end retrieval-stage
  latency by roughly two-thirds.
\item \textbf{Online results and a per-source gap analysis} (Section~\ref{sec:ab}).
  In online A/B tests, \grp{} as a retrieval source ranks near the top on
  completion and sharing and above the all-source average on almost every
  engagement axis; RL lifts its view-time contribution over the supervised model,
  and a larger decode budget lifts view time over production on its own; and
  replacing lower-performing sources with \grp{} yields significant gains in
  views, view time, and shares on the surface with neutral guardrails.
\end{enumerate}

The remainder of the report is organized as follows. Section~\ref{sec:progressive}
lays out the progressive path and its measurement protocol.
Section~\ref{sec:pretrain} presents the architecture and pre-training results,
Section~\ref{sec:posttrain} the post-training method, and
Section~\ref{sec:deploy} the serving stack. Section~\ref{sec:ab} reports the
online A/B test, and Section~\ref{sec:conclusion} discusses limitations and the
next steps toward full replacement.

\section{A Progressive Path to End-to-End Generative Recommendation}
\label{sec:progressive}

We pursue E2E generative recommendation through a staged migration from the
production cascade to a unified model. We therefore stage the migration so that every step is an
independently measurable, shippable change, and so that the comparison against the
incumbent system is always made on the incumbent's own terms.

\subsection{Three phases}

\paragraph{Phase 1: deploy as a retrieval source and benchmark.}
The generative model enters the production funnel as \emph{one more retrieval
source}, alongside many incumbent retrieval sources. It receives a fixed retrieval quota and its candidates flow
through the unchanged downstream rankers. Because every source is measured on the
same recommendation metrics for the same users, this immediately tells us which incumbent
sources the generator already outperforms and where it falls short---and any
inferiority pinpoints the specific bottleneck that must be solved before full
replacement is plausible.

\paragraph{Phase 2: replace weaker sources and grow the quota.}
Sources that the generator dominates on the axes that matter are retired and their
quota is reassigned to the generator. As the model improves, its quota grows until
only a minimal set of auxiliary sources is needed to preserve properties the
generator does not yet provide (e.g., exploration or freshness).
This is the phase the present report reaches (Section~\ref{sec:ab}).

\paragraph{Phase 3: bypass the rankers.}
Because \grp{} emits its own calibrated candidate scores, a growing fraction of
its candidates can bypass first the early-stage and eventually the late-stage
rankers. When the bypassed fraction reaches a threshold at which the rankers no
longer change the served slate, they become obsolete and the migration is complete.
The serving stack already exposes the control needed for this phase
(Section~\ref{sec:serving-arch}), and the online experiments of
Section~\ref{sec:ab} already exercise its first step by letting a portion of
\grp{} candidates bypass early ranking.

\subsection{Measurement protocol}
\label{sec:protocol}

The protocol has three tiers, all computed on the treatment arm.

\begin{itemize}[leftmargin=1.6em]
\item \textbf{Surface-level evaluation metrics} on the short-video surface: views, view
  time, and shares. Views and view time describe content consumption; shares measure use of the
  sharing feature. These metrics are interpreted together.
\item \textbf{Topline guardrails}, such as DAUs and app-level engagement,
  to check for adverse platform-level effects.
\item \textbf{Per-source performance}, which is what makes the progressive path
  navigable: \emph{source rate} (the share of served views attributed to a
  source), \emph{ranking pass rate} (the share of a source's candidates that
  survive the ranking stages), and the source's \emph{engagement profile}
  (completion ratio, skip rate, average watch time, favorite rate, and send rate).
\end{itemize}

\section{Model Architecture and Pre-training}
\label{sec:pretrain}

\subsection{Overview}
\label{sec:overview}

\begin{figure}[!t]
\centering
\includegraphics[width=\textwidth]{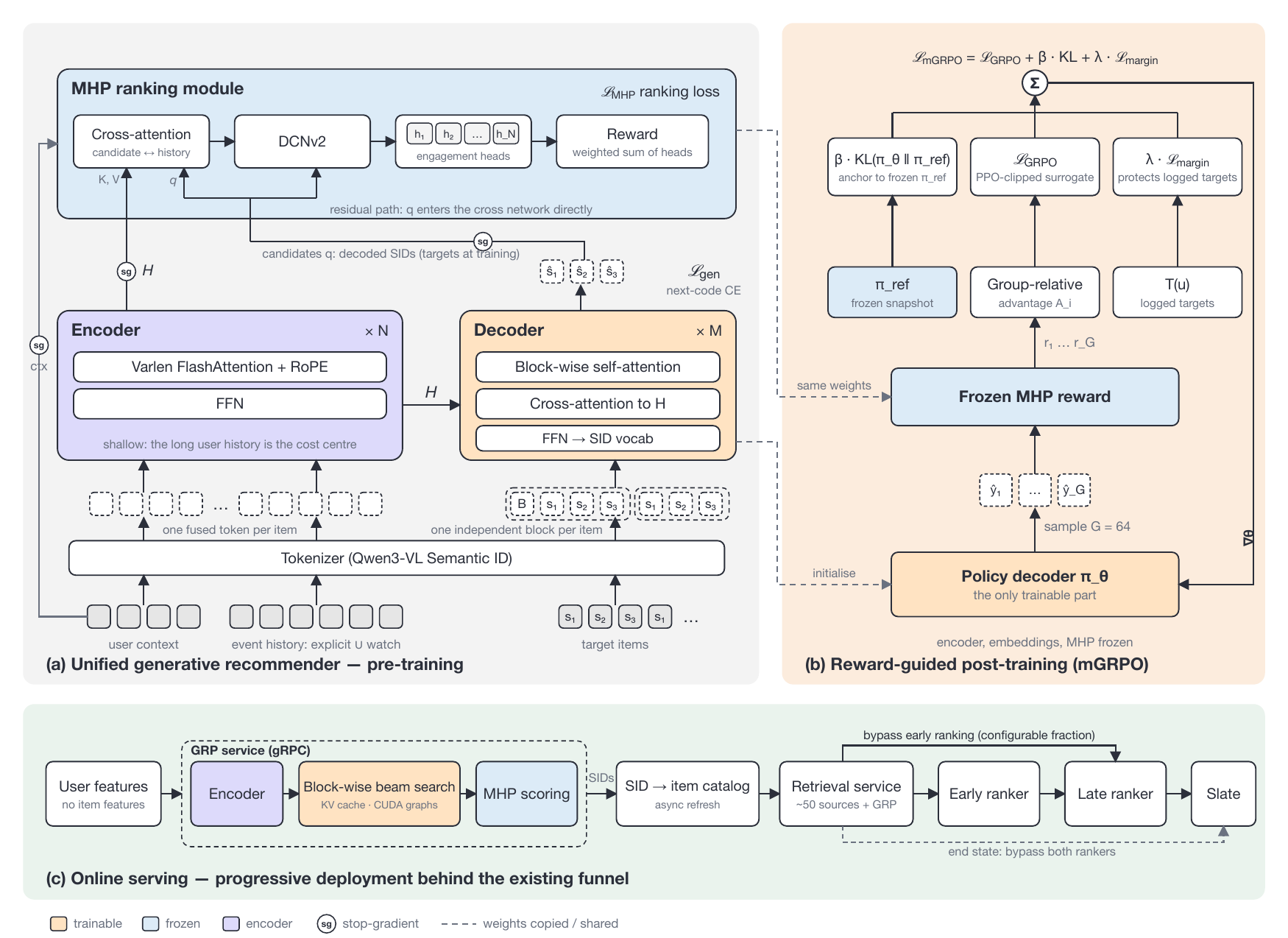}
\caption{Overall architecture of \grp{}. \textbf{(a) Pre-training.} User context
and the merged explicit-plus-watch event history are tokenized into Qwen3-VL
Semantic IDs and fused into one encoder token per item; target items form
independent per-item code blocks for the decoder. The encoder (shallow) produces
$H$; the decoder (deep) cross-attends $H$ and generates the next item's Semantic
ID block-wise (generation loss $\mathcal{L}_{\text{gen}}$). The \emph{MHP ranking
module} receives $H$, the user context, and the candidate query $q$ through
stop-gradient (sg) paths: $q$ cross-attends the history and also enters the DCNv2
cross network directly (the residual path); per-head predictions give the ranking
loss $\mathcal{L}_{\text{MHP}}$ and a weighted-sum reward. At training time the
candidates are the logged targets; at serving time they are the decoded Semantic
IDs. \textbf{(b) Post-training.} Only the decoder trains as the policy
$\pi_\theta$: it samples $G$ Semantic IDs per user, the frozen MHP (same weights
as in (a)) scores them, and the group-relative advantage feeds the PPO-clipped
surrogate; a KL term anchors the policy to the frozen snapshot $\pi_{\text{ref}}$
and the one-sided, reference-anchored margin protects the logged targets
$\mathcal{T}(u)$ (\mgrpo{}, Section~\ref{sec:mgrpo}). \textbf{(c) Online serving.}
The inference path runs the encoder, block-wise beam search (KV cache, CUDA graphs),
and MHP scoring per request and returns Semantic IDs with scores; an
asynchronously refreshed catalog maps them to items, which enter the existing
retrieval service as an additional source. A configurable fraction of
\grp{} candidates bypasses early ranking on the model's own scores; the end state
bypasses both rankers (Section~\ref{sec:progressive}).}
\label{fig:arch}
\end{figure}

\grp{} is a single model with two heads over a shared encoder--decoder trunk
(Figure~\ref{fig:arch}). The \textbf{encoder} ingests a long, heterogeneous
user-context and event-history sequence and produces contextual hidden states
$H$. The \textbf{decoder} generates the Semantic ID of the next item---this is the
\emph{retrieval} output. The \textbf{MHP ranking module} takes the generated (or
ground-truth, at training time) Semantic-ID candidates, cross-attends them against
$H$ and the user context, and emits a vector of calibrated engagement
predictions---this is the \emph{ranking} output. In post-training, a fixed
weighted mixture of the frozen MHP heads becomes the \emph{reward model} for
reinforcement-learning fine-tuning of the decoder.

Training proceeds in two phases. \textbf{Pre-training} (this section) jointly
learns generation (next-Semantic-ID cross-entropy) and ranking (per-engagement
classification and regression), with the ranking module's inputs detached so it
cannot perturb the generation trunk. \textbf{Post-training}
(Section~\ref{sec:posttrain}) freezes the encoder, the embedding tables, the
normalization statistics, and the MHP module, and optimizes only the decoder
against the frozen reward.

\subsection{Tokenization: Qwen3-VL Semantic IDs}
\label{sec:tokenization}

Items are represented by \textbf{Semantic IDs (SIDs)}: each item is quantized into
a short sequence of hierarchical discrete codes by a residual-quantization
variational autoencoder (RQ-VAE)~\citep{rqvae2021,tiger2023}. The production
tokenizer is the \textbf{Qwen3-VL SID}~\citep{qwen3vl2025}: a Qwen3-VL
vision--language embedding (natively video-aware, Matryoshka-truncated for a large
storage/compute saving) is residual-quantized into three hierarchy levels of
decreasing codebook size. Beyond the usual reconstruction and commitment terms, the
tokenizer adds an auxiliary \emph{co-engagement contrastive} objective---items
that are co-engaged by similar users are pulled together---which improves codebook
utilization and item uniqueness over the prior tokenizer. Codebook collapse is
mitigated with a straight-through estimator and post-quantization normalization.

All SID codes share a \textbf{single embedding table}: levels are laid out
contiguously by cumulative offset and a separator/BOS token is appended, so the
vocabulary size is the sum of the per-level codebook sizes plus one. Each item
therefore occupies one \textbf{fixed-length block} of codes (one per hierarchy
level) followed by a separator, in both the history and the target---a structure
the decoder exploits for block-wise independence (Section~\ref{sec:generation}).
Multiple SID versions coexist in the feature store as version-tagged,
variable-width records, so the generative vocabulary can be upgraded without
re-ingesting content.

\subsection{Generation module (encoder--decoder)}
\label{sec:generation}

The trunk is a T5-style encoder--decoder~\citep{t5_2020}. The decoder is
intentionally \textbf{deeper than the encoder} (a $1{:}2$ encoder-to-decoder
layer ratio in the reported configuration); Section~\ref{sec:encdec} shows this decoder-heavy
allocation is a Pareto improvement in both accuracy and throughput. All attention
modules are replaced in place by a custom FlashAttention-based implementation.

\paragraph{User-context tokenization.}
The encoder input is built from two parts. The \emph{user-context} vector
concatenates a universal user-model embedding with static dense and sparse-feature
embeddings, and is projected into a few context tokens of width $d_{\text{model}}$.
The \emph{event history} is grouped into two sequences: an
\emph{explicit-engagement} group (high-intent actions such as favoriting or
resharing a video, each carrying a learned action-type embedding) and a
\emph{watch} group (passive views). Each group's features are fused with its SID
embeddings, projected per group, and concatenated.

\paragraph{Item-level history encoding.}
We encode the history at the \emph{item} level: the hierarchy codes and side
features of each history item are fused into a \emph{single} encoder token, so an
encoder token corresponds to an item while a decoder token corresponds to one SID
code. Let $B_{\mathrm{ref}}$ denote the reference history budget, defined as the number of
history items accommodated by the reported item-level configuration.
SID-level encoding used four encoder positions per item and accommodated
$B_{\mathrm{ref}}/4$ history items; item-level fusion accommodates $B_{\mathrm{ref}}$ items within
the same encoder-token budget. Section~\ref{sec:itemenc} shows
this is effectively free on quality.

\paragraph{Per-user event sampling.}
Histories are capped at a fixed total budget $B$ of events, split between the two
groups so that scarce high-intent events are not crowded out by abundant passive
views---without such a rule, watch events alone fill the budget. The split
cannot be a fixed per-type quota, however: the mix of event types varies
substantially from user to user, so any fixed allocation leaves slots empty for
users who rarely take a given action while truncating the history of users who
take it often, and the total budget goes underused. We therefore allocate slots
per user, from that user's own event counts. The explicit-engagement group is given a protected budget $B_e < B$, allocated across
its event types by a \emph{sublinear} rule with a per-type floor: for event type
$t$ with raw count $n_t$,
\begin{equation}
s_t \;=\; \max\!\Big(f,\ \Big\lfloor B_e \cdot
\frac{n_t^{\,p}}{\sum_{t'} n_{t'}^{\,p}} \Big\rfloor\Big),
\qquad p \in (0,1),
\label{eq:sampling}
\end{equation}
and the watch group elastically fills the remaining $B - \sum_t s_t$ slots. The
sublinear exponent $p$ and floor $f$ keep rare but valuable actions represented
while a high-volume type cannot monopolize the budget. Within each group events
are kept most-recent-first. Section~\ref{sec:events} quantifies the effect.

\paragraph{Positional encoding.}
We use rotary position embeddings (RoPE)~\citep{rope2021}, applied \textbf{only to
encoder self-attention}. Decoder self-attention operates within per-item blocks
where local ordering suffices, and cross-attention spans two different semantic
spaces (history vs.\ SID), so RoPE is deliberately disabled for both. The
relative-position bias of the base transformer is discarded.

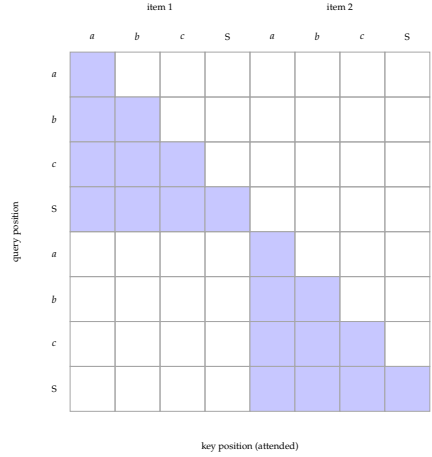
\begin{figure}[t]
\centering
\resizebox{0.34\textwidth}{!}{%
\begin{tikzpicture}[font=\tiny]
\foreach \i in {0,...,7}{
  \foreach \j in {0,...,7}{
    \pgfmathtruncatemacro{\bi}{int(\i/4)} \pgfmathtruncatemacro{\bj}{int(\j/4)}
    \ifnum\bi=\bj \ifnum\j>\i\else \fill[blue!22] (\j,-\i) rectangle ++(1,-1); \fi\fi
    \draw[black!35] (\j,-\i) rectangle ++(1,-1);
  }
}
\foreach \j/\l in {0/$a$,1/$b$,2/$c$,3/S,4/$a$,5/$b$,6/$c$,7/S}{
  \node[font=\tiny] at (\j+0.5,0.35) {\l};
  \node[font=\tiny] at (-0.35,-\j-0.5) {\l};}
\node[font=\tiny] at (2,1.0) {item 1}; \node[font=\tiny] at (6,1.0) {item 2};
\node[font=\tiny] at (4,-8.8) {key position (attended)};
\node[font=\tiny, rotate=90] at (-1.2,-4) {query position};
\end{tikzpicture}}
\caption{The decoder's block-wise self-attention mask: causal \emph{within} each
item's code block (shaded lower-triangle) and blocked \emph{across} blocks, so the
target items are decoded independently.}
\label{fig:blockwise}
\end{figure}

\paragraph{Block-wise decoding.}
A defining design choice is to decode the target items in a request
\textbf{independently}. We represent the recommendation slate as a set of
next-item candidates and impose no causal dependency from one predicted item to
the next (Figure~\ref{fig:blockwise}). We enforce this in
decoder self-attention by partitioning the target sequence into one block per
candidate item and attending \textbf{independently within each block}. This keeps
the decoding of each target item independent and, both during training and during
beam search, avoids broadcasting the encoder states across the full set of targets
(or beams) until they are actually needed---a naive broadcast bloats activation
memory and readily exhausts GPU memory at production sequence lengths. The net
effect is that generation cost scales with the number of targets rather than their
cross-product, and the same forward pass emits a whole slate of independent
candidates.

\paragraph{Variable-length attention.}
Encoder self-attention and decoder cross-attention use variable-length
FlashAttention~\citep{flashattention2022} kernels with input packing, so padded
positions in the highly variable-length user histories incur no compute; decoder
self-attention is causal within blocks.

\paragraph{Generation loss.}
Generation is trained with token-level cross-entropy over the shared code
vocabulary under teacher forcing:
\begin{equation}
\mathcal{L}_{\text{gen}}
= -\frac{1}{|\mathcal{M}|}\sum_{t\in\mathcal{M}}
   \log p_\theta\!\left(y_t \mid y_{<t}^{\text{blk}}, H\right),
\label{eq:gen}
\end{equation}
where $H$ are the encoder hidden states, $y_{<t}^{\text{blk}}$ the
teacher-forced codes preceding position $t$ within its target block, and
$\mathcal{M}$ the set of content-code positions---every per-item separator
position (and padding) is masked out. The objective is thus next-code prediction
over the hierarchy levels of each target item.

\subsection{Ranking module (MHP)}
\label{sec:mhp}

The ranking module is a \textbf{multi-head prediction (MHP)} network that turns
candidates into calibrated engagement scores while remaining \emph{serving-safe}:
it never consumes decoder outputs, so it can score arbitrary candidates at serving
time, including candidates that were not generated by the decoder.

\paragraph{Detached input.}
All three inputs to the MHP module---the candidate SID embeddings, the encoder
hidden states, and the user-context features---are \textbf{detached
(stop-gradient)} before entering the module. Consequently the ranking loss trains
only the MHP's own parameters and never backpropagates into the encoder, decoder,
or shared SID table. This is the mechanism that lets us train ranking
\emph{jointly} with generation without the ranking objective distorting the
generative representation.

\paragraph{Architecture.}
For each candidate item, its hierarchy SID embeddings are pooled into a
\emph{candidate query} $q$. The query cross-attends (multi-head attention) against
the encoded user history to form a \emph{candidate-aware user vector}; this vector,
the broadcast user-context features, \emph{and the candidate query itself} are
concatenated and passed through a \textbf{DCNv2} cross network~\citep{dcnv2_2021}
for explicit feature interaction. Each engagement label then gets its own small MLP
prediction head. The direct (residual) path from $q$ into the cross network is
essential: Section~\ref{sec:reward-fix} shows that without it the module learns to
ignore the candidate altogether. The cross-network and head widths are matched to
the production second-stage ranker, each head's output bias is initialized from
its label's base rate, and classification logits are clamped for numerical
stability.

\paragraph{Ranking loss.}
Each head is trained against its own engagement label, masked to valid target
positions:
\begin{equation}
\mathcal{L}_{\text{MHP}}
= \sum_{h\in\mathcal{C}} \mathrm{BCE}\!\left(\hat{y}_h, y_h\right)
+ \sum_{h\in\mathcal{R}} \mathrm{MSE}\!\left(\hat{y}_h, y_h\right),
\label{eq:mhp}
\end{equation}
where $\mathcal{C}$ are binary classification heads over positive engagements
(e.g., completing, favoriting, or resharing an item) and negative signals (e.g.,
skipping or dismissing it), and $\mathcal{R}$ is a regression head for (log) watch
time. Each head is weighted equally. Calibration of these heads is a first-class
concern: because the frozen module later serves as the RL reward, a miscalibrated
head directly distorts that reward, so we monitor head calibration continuously.

\subsection{Training objective and optimizers}
\label{sec:finalloss}

The total pre-training objective is a plain sum of the generation loss and every
ranking-head loss:
\begin{equation}
\mathcal{L}_{\text{pretrain}}
= \mathcal{L}_{\text{gen}} + \mathcal{L}_{\text{MHP}}.
\label{eq:total}
\end{equation}
There is no learned or uncertainty-based task weighting. Because the MHP inputs are
detached, the head losses do not propagate into the generation trunk---the sum is
a bookkeeping convenience, and the two objectives are decoupled at the gradient
level while sharing one forward pass and one set of features.

Dense parameters are optimized by a single hybrid optimizer. Weight matrices
($\ge 2$-D) are updated by \textbf{Muon}~\citep{muon2024}---momentum followed by a
few steps of Newton--Schulz orthogonalization of the update---while 1-D parameters
(biases, norms) are updated by \textbf{Lamb}~\citep{lamb2020}. Sharded sparse
embeddings are excluded and handled by a row-wise adaptive optimizer in the
embedding path. Both branches use \textbf{fused, batched kernels} (one stacked
orthogonalization per parameter-shape group), which improve training throughput by
more than 30\% and are what make Muon practical at this scale. Learning rates
follow square-root scaling with the global batch size $B_g$,
$\eta = \eta_0 \sqrt{B_g / B_0}$, with a per-branch base rate $\eta_0$ at
reference batch $B_0$, under linear warmup followed by cosine annealing. Training
uses bf16 mixed precision.

\subsection{Empirical results}
\label{sec:pretrain-results}

Unless noted, offline metrics are computed on held-out user sessions from the
production short-video surface, evaluated on a future date no model trained on,
over a large session population. We report \textbf{HR@$K$} (hit rate: the
ground-truth next item's Semantic ID appears among the top-$K$ generated SIDs),
\textbf{Recall@$K$} (the fraction of a user's held-out engaged items retrieved
within the top-$K$ candidates), and \textbf{Reward Recall@$K$} (recall weighted by
each retrieved item's reward under the specified reward function; this metric
measures alignment with the post-training objective).
Each study below is a one-factor comparison against its own baseline.

\subsubsection{Scaling the user-history length with item-level fusion}
\label{sec:itemenc}

Encoding each history item as a single fused token rather than one token per SID
code shortens the encoder sequence $4\times$ at a fixed number of history items.
Table~\ref{tab:itemenc} isolates this factor at a matched encoder-token length:
item-level encoding fits $4\times$ the history ($B_{\mathrm{ref}}$ vs.\ $B_{\mathrm{ref}}/4$ items) at essentially
unchanged throughput and a slightly higher hit rate (HR@10 $+1.7\%$, HR@20
$+0.9\%$). Extending the history a further $2\times$ (to $2B_{\mathrm{ref}}$) stops
helping retrieval and halves throughput; $B_{\mathrm{ref}}$ is the sweet spot at this model
capacity, a result later reconfirmed on independent seven-day data
(Section~\ref{sec:events}).

\begin{table}[t]
\centering
\caption{History encoding, one-factor comparison on a common held-out evaluation
(a $2{:}1$ encoder-to-decoder layer ratio; throughput in training steps/s). At a matched
encoder-token length, item-level encoding fits $4\times$ the history at essentially
the same throughput and a slightly higher hit rate; extending to $8\times$
($2B_{\mathrm{ref}}$ items) does not improve retrieval and halves throughput.
History budgets are reported relative to $B_{\mathrm{ref}}$.}
\label{tab:itemenc}
\small
\begin{tabular}{lccccc}
\toprule
Representation & Relative history budget & HR@5 & HR@10 & HR@20 & Steps/s \\
\midrule
SID-level          & $0.25\times$ & 0.1207 & 0.1734 & 0.2272 & 0.58 \\
Item-level fusion  & $1\times$ & 0.1217 & 0.1764 & 0.2293 & 0.57 \\
Item-level fusion  & $2\times$ & 0.1214 & 0.1751 & 0.2304 & 0.32 \\
\midrule
Change, $1\times$ vs.\ SID-level & $+300\%$ & $+0.83\%$ & $+1.73\%$ & $+0.92\%$ & $-1.72\%$ \\
\bottomrule
\end{tabular}
\end{table}

\subsubsection{Rebalancing dense capacity toward the decoder}
\label{sec:encdec}

Item-level fusion makes history encoding cheap, which raises the question of where
dense capacity is best spent. Because the encoder sequence (the long user history)
is far longer than the decoder sequence (a short SID target), the encoder runs full
attention over the cost center while the decoder runs over a cheap target. Shifting layers from the encoder to the decoder is therefore a
\textbf{Pareto win} (Table~\ref{tab:encdec}): as the encoder/decoder layer split
moves $2{:}1 \rightarrow 1{:}1 \rightarrow 1{:}2$ at a fixed total depth,
HR@5/10/20 and validation loss improve monotonically \emph{and} throughput rises
by 21\%. The reported model therefore uses a $1{:}2$ encoder-to-decoder ratio.

\begin{table}[t]
\centering
\caption{Encoder/decoder rebalancing at a fixed $B_{\mathrm{ref}}$-item history and total depth.
Moving layers from the encoder to the decoder improves retrieval (HR@$K$) and
training throughput simultaneously.}
\label{tab:encdec}
\small
\begin{tabular}{lcccc}
\toprule
Encoder : decoder layer ratio & HR@5 & HR@10 & HR@20 & Steps/s \\
\midrule
$2{:}1$ & 0.1217 & 0.1764 & 0.2293 & 0.57 \\
$1{:}1$ & 0.1232 & 0.1771 & 0.2318 & 0.63 \\
$1{:}2$ & 0.1241 & 0.1783 & 0.2332 & 0.69 \\
\midrule
Change, $1{:}2$ vs.\ $2{:}1$ & $+1.97\%$ & $+1.08\%$ & $+1.70\%$ & $+21.05\%$ \\
\bottomrule
\end{tabular}
\end{table}

\subsubsection{Preserving sparse explicit actions in longer histories}
\label{sec:events}

A longer history is only useful if it contains the right events. Three
observations about the pre-trained model drive the sampling policy of
Equation~\eqref{eq:sampling}. First, retrieval quality counterintuitively
\emph{falls} as a user's history grows: the longest-history segment is the hardest, because long, diverse
histories make the next item less predictable and a plain recency cap discards earlier interactions. Second, the generator over-concentrates on
popular head content, whereas explicitly engaged items are markedly more niche.
Third, anchoring the history to the passive-watch stream lost many scarce
high-intent events to that stream's truncation cap.

We therefore merge the explicit-engagement events (favorites, shares, replies,
and other high-intent actions) into one ordered stream with reserved capacity, so
abundant watches cannot crowd them out, and let the watch history fill the
remaining budget.
Table~\ref{tab:events} is a clean one-factor ladder on seven-day training data
with filter-free held-out targets. Item-level encoding \emph{alone} is a small
negative here (the history is longer but still watch-dominated); adding the merged explicit stream is the step that moves reward-recall; and budget scaling past $B_{\mathrm{ref}}$ is flat to down ($2B_{\mathrm{ref}}$ is
slightly worse), which we attribute to insufficient model capacity for longer
sequences rather than to truncation---we verified that 87\% of the window is
utilized and that the most recent events are always kept. The merge design
mainly benefits reward-weighted recall: against the item-level baseline it
improves reward-recall@20 by 20\% with a smaller gain in raw recall@20 ($+7\%$).
The reported configuration reserves 25\% of its total history budget $B_{\mathrm{ref}}$ for explicit events.

\begin{table}[t]
\centering
\caption{Event-selection ladder (seven-day training, one change per row). Merged-stream rows show the explicit-event
fraction of the total budget, followed by the total budget relative to $B_{\mathrm{ref}}$. Merging explicit events into a protected stream
is the only step that moves reward-recall materially.}
\label{tab:events}
\small
\begin{tabular}{lccc}
\toprule
Configuration & Reward Recall@20 & Recall@20 & HR@20 \\
\midrule
(a) SID-level history, recency window & 0.0252 & 0.0521 & 0.1356 \\
(b) Item-level history, no merged explicit stream & 0.0245 & 0.0502 & 0.1327 \\
(c) Merged explicit, 50\%; $0.25\times$ & 0.0291 & 0.0543 & 0.1487 \\
(d) Merged explicit, 25\%; $1\times$ & 0.0294 & 0.0537 & 0.1481 \\
(e) Merged explicit, 12.5\%; $2\times$ & 0.0290 & 0.0519 & 0.1462 \\
\midrule
Change, (d) vs.\ (b) & $+20.0\%$ & $+7.0\%$ & $+11.6\%$ \\
\bottomrule
\end{tabular}
\end{table}

\subsubsection{A higher-resolution Semantic ID improves retrieval}
\label{sec:sid}

The tokenizer is one of the highest-leverage design choices, because the
generative vocabulary \emph{is} the SID space: the model can only retrieve items
whose Semantic ID it can generate, so a tokenizer that packs more distinct,
semantically coherent items into the codebook directly raises the retrieval
ceiling. We therefore measure a tokenizer by the downstream model's retrieval
quality, not by intrinsic codebook statistics alone.
Table~\ref{tab:sid} compares two tokenizers under a matched setup at a
common codebook budget: the earlier \emph{MM-SID} (a uniform, lower-resolution
multimodal codebook) and the \emph{Qwen3-VL SID} (a higher-resolution,
non-uniform codebook). Swapping MM-SID for the Qwen3-VL SID improves recall@20 by
11.9\%, reward-recall@20 by 10.3\%, and HR@20 by 9.9\%. Because the Qwen3-VL SID
has a larger code-tuple space (an exact code-tuple match is a strictly harder
target), these metrics are, if anything, biased against it and \emph{understate}
the gain in retrievable coverage. The mechanism is codebook quality: the natively
video-aware embedding and the co-engagement contrastive objective spread items far
more evenly across the codebook and roughly double item uniqueness over the prior
tokenizer.

\begin{table}[t]
\centering
\caption{Downstream retrieval quality by Semantic-ID version at a matched model
size, codebook budget, and train/held-out evaluation (a one-factor SID swap).}
\label{tab:sid}
\small
\begin{tabular}{lccc}
\toprule
Semantic-ID version & Recall@20 & Reward Recall@20 & HR@20 \\
\midrule
MM-SID       & 0.0529 & 0.0398 & 0.1563 \\
Qwen3-VL SID & 0.0592 & 0.0439 & 0.1717 \\
\midrule
Change       & $+11.9\%$ & $+10.3\%$ & $+9.9\%$ \\
\bottomrule
\end{tabular}
\end{table}

\subsubsection{Freshness dominates out-of-window}
\label{sec:freshness}

\begin{figure}[t]
\centering
\begin{tikzpicture}
\begin{axis}[
  width=0.62\textwidth, height=5.2cm,
  xlabel={Days past training cutoff}, ylabel={Recall (rel.\ to in-window)},
  xmin=-0.2, xmax=4.2, ymin=0, ymax=1.08, xtick={0,1,2,3,4}, ytick={0,0.5,1.0},
  tick label style={font=\footnotesize}, label style={font=\footnotesize},
  legend style={at={(0.97,0.83)}, anchor=north east, font=\scriptsize},
  every axis plot/.append style={thick, mark size=2pt}]
\addplot[blue, mark=*] coordinates {(0,1.0) (1,0.686) (2,0.388) (3,0.286) (4,0.237)};
\addplot[orange, mark=square*, dashed] coordinates {(0,1.0) (1,0.99) (2,0.98) (3,0.97) (4,0.97)};
\legend{exact-item recall, coarse-cluster recall}
\end{axis}
\end{tikzpicture}
\caption{Freshness decay of a frozen pre-trained model (recall normalized to its
in-window value). Exact-item recall collapses within days as fresh items go
out-of-vocabulary, while coarse-cluster recall is essentially flat, consistent with continued
prediction of broad content preferences as exact-item coverage declines. We
address catalog freshness through continuous incremental retraining; post-training
adjusts candidate probabilities using the specified reward function.}
\label{fig:freshness}
\end{figure}
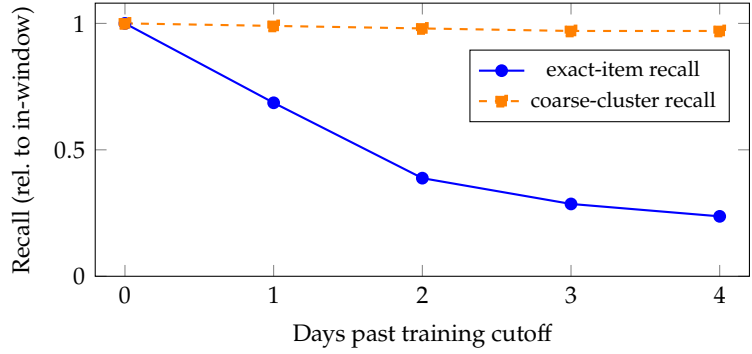

The pre-trained model's exact-item retrieval quality \textbf{declines sharply
within a few days of the training cutoff} (Figure~\ref{fig:freshness}): the catalog turns over almost completely
day-over-day, so the target set becomes overwhelmingly novel and fresh items fall
out of vocabulary. Crucially, \emph{coarse}-cluster recall stays essentially flat
over the same window---the \emph{content} a user wants is still predictable; only
the \emph{exact fresh item} is unreachable. We address this freshness limitation with \textbf{continuous incremental
retraining} to incorporate newly available items. This
limitation also explains why the post-training scope (Section~\ref{sec:posttrain}) is confined to
reshaping the in-window operating point: RL can redistribute generation
probability among items the model can represent, but it cannot manufacture
representations for post-cutoff items.

\section{Post-training: Reward-Guided Alignment}
\label{sec:posttrain}

Pre-training optimizes the decoder for next-code likelihood over ground-truth
Semantic IDs, learning from logged user behavior without weighting outcomes by a specified
reward. Different interaction types can receive different weights under that
reward, which next-code likelihood alone does not express. We close this gap
with an on-policy Group-Relative Policy Optimization (GRPO)~\citep{grpo2024}
post-training stage that treats the pre-trained recommender as a policy and its
own ranking module as the reward. We identified two failure modes:
Sections~\ref{sec:reward-fix} and~\ref{sec:mgrpo} proceed from
each problem to the intervention and the measured outcome.

\subsection{GRPO against a frozen in-model reward}
\label{sec:grpo}

\paragraph{Frozen trunk.}
We freeze the encoder, the sparse and Semantic-ID embedding tables, the
normalization statistics, and the entire MHP module, and train \emph{only the
decoder} (the decoder stack, its input LayerNorm, and the output projection).
Freezing the trunk keeps the MHP reward a \textbf{stationary} function of
$(\text{prompt}, \text{sample})$: the environment signal does not drift while the
policy learns, which stabilizes the group-relative advantage and avoids the
reward-hacking dynamics that a co-adapting reward model would introduce. Decoder
dropout is disabled so that policy and reference log-probabilities are directly
comparable.

\paragraph{Sampling and reward.}
For each prompt $u$ (an encoded user context), the current policy draws a group
$\mathcal{G}(u)$ of $G$ candidate Semantic-ID sequences by temperature sampling
(no beam search): the encoder output is expanded $G$-fold and the decoder samples
one code per hierarchy level, recording the per-token behavior log-probabilities.
Each sampled sequence is scored by the \textbf{frozen MHP reward model} through the
same serving-safe path used in pre-training, and a scalar reward is a fixed
weighted mixture over heads,
\begin{equation}
r \;=\; \sum_{h} w_h\, \phi_h(\hat{y}_h),
\label{eq:reward}
\end{equation}
where $\phi_h$ maps a head output to its served value (a sigmoid for
classification heads, an exponential for the log-watch-time head) and the weights
$w_h$ specify the contribution of each prediction head. Reward scoring runs without
gradients: the MHP is a frozen environment signal. For the reported RL comparison in Section~\ref{sec:ab}, the view-time prediction
head is the sole reward.

\paragraph{GRPO objective.}
Within each group the advantage is the standardized reward, and the policy loss is
the PPO-clipped surrogate~\citep{ppo2017} at the item level with an exact
analytic per-token KL penalty to a frozen reference policy:
\begin{equation}
A_i = \frac{r_i - \mu_{\mathcal{G}}}{\sigma_{\mathcal{G}} + \delta},
\qquad
\rho_i(\theta) = \frac{P_\theta(\hat{y}_i \mid u)}{P_{\theta_{\text{old}}}(\hat{y}_i \mid u)},
\label{eq:adv}
\end{equation}
\begin{equation}
\mathcal{L}_{\text{GRPO}}(\theta)
= -\frac{1}{G}\sum_{i=1}^{G}
  \min\!\big(\rho_i A_i,\ \operatorname{clip}(\rho_i, 1{-}\epsilon, 1{+}\epsilon)\,A_i\big)
+ \beta\, \overline{D}_{\mathrm{KL}}\!\left(\pi_\theta \,\|\, \pi_{\text{ref}}\right),
\label{eq:grpo}
\end{equation}
where $P_\theta(\hat y \mid u)$ is the item probability (the product of
teacher-forced code probabilities over the content positions), $\epsilon$ is the
clip range, $\beta$ the KL coefficient, and $\overline{D}_{\mathrm{KL}}$ the KL
between the policy and reference next-code distributions over the full
Semantic-ID vocabulary, averaged over valid token positions. The reference policy
$\pi_{\text{ref}}$ is a detached snapshot of the decoder weights, captured once at
the start of post-training and evaluated with a functional forward; because the
encoder, embeddings, and MHP are frozen and shared, only the decoder is duplicated
rather than the full model. Post-training initializes from the pre-trained checkpoint, resets
the learning-rate schedule, reduces the per-device batch size to absorb the
$G$-fold rollout, and reuses the same optimizers at reduced, batch-scaled learning
rates.

\subsection{Making the reward discriminate between candidates}
\label{sec:reward-fix}

\paragraph{Problem.}
Our first GRPO runs were noisy and learned little. Instrumenting the reward
revealed why: rewards were nearly \emph{candidate-indifferent} within a user. Across
the candidates sampled for a single user request, the mixed reward varied by only $0.005$--$0.16\%$ of its
magnitude (individual heads by $0.3$--$2.3\%$), so the standardized advantages of
Equation~\eqref{eq:adv} looked healthy ($\pm 1$) while carrying little
ranking information per group. Tracing the candidate signal stage by stage through
the module localized the loss to the cross-attention layer: its attention pattern
had become effectively independent of the candidate query (the learned query bias
dominated the query projection by three orders of magnitude), so the layer acted
as a constant key-selector over the history and discarded essentially all of the
candidate's information in one step. The module had learned $P(\text{engage}\mid
\text{user})$ rather than $P(\text{engage}\mid\text{user},\text{item})$---a
degenerate solution that pre-training never penalized, because the MHP only ever
saw the logged target and could minimize its loss from the user context alone.

\paragraph{Intervention.}
We redesigned the MHP input path (Section~\ref{sec:mhp}) so that the candidate
query is a \emph{direct} input to the cross network, i.e., candidates pass through
the module along a residual path that no attention layer can zero out; adjusted the
cross-network and prediction-head capacity;
initialized each head from its base-rate prior; and clamped classification logits.

\paragraph{Result.}
On identical evaluation sets (Table~\ref{tab:mhp}), three of the
four dense heads gained $0.11$--$0.13$ AUROC and all three reported sparse heads gained
$97$--$184\%$ AUPRC; the skip head was the only regression ($-0.018$ AUROC), which may be within noise.
Recall@20 ($+0.0004$) and reward-recall@20 ($+0.0001$) were unchanged, confirming
that the MHP change did not leak into decoding---as the detached design intends.
On the redesigned module the within-user reward spread rises from $<0.2\%$ to a
median of over 200\% of the reward magnitude, which is what makes the group-relative
advantage informative.

\begin{table}[t]
\centering
\caption{Ranking-module redesign, evaluated on identical held-out
sets. Dense (frequent) heads are reported by AUROC and sparse (rare) heads by
AUPRC. Watch$\ge$10s and Watch$\ge$p75 are watch-time thresholds (absolute, and
relative to the item's duration distribution).}
\label{tab:mhp}
\small
\begin{tabular}{llccc}
\toprule
Head & Metric & Before & After & Change \\
\midrule
Watch $\ge$ 10\,s          & AUROC & 0.6628 & 0.7759 & $+17.1\%$ \\
Skip                       & AUROC & 0.6281 & 0.6102 & $-2.8\%$ \\
Complete                   & AUROC & 0.6294 & 0.7609 & $+20.9\%$ \\
Watch $\ge$ p75            & AUROC & 0.7046 & 0.8296 & $+17.7\%$ \\
\midrule
Share                      & AUPRC & 0.0213 & 0.0457 & $+114.6\%$ \\
Favorite                   & AUPRC & 0.0301 & 0.0592 & $+96.7\%$ \\
Reply                      & AUPRC & 0.0123 & 0.0349 & $+183.7\%$ \\
\bottomrule
\end{tabular}
\end{table}

\subsection{\mgrpo{}: stopping reward optimization from collapsing recall}
\label{sec:mgrpo}

\paragraph{Problem.}
Even with a discriminative reward, vanilla GRPO frequently produced no gain or a
net loss. The mechanism is a cascade: the pre-trained recommender's recall of
logged targets is modest to begin with, so most sampled groups contain no target
and RL exploration is driven by reward alone; the policy drifts toward
high-reward regions that the base model never validated against real behavior; and
as it drifts, its recall of real targets erodes further, which removes the very
signal that would anchor it. The KL penalty of Equation~\eqref{eq:grpo} slows this
drift but does not target it, because KL is agnostic to \emph{which} probability
mass moves. In our experiments, the standard remedies---tuning the KL coefficient
$\beta$, or adding a supervised likelihood loss on the logged targets---produced
either negative or only small improvements, which led us to the targeted
alternative below.

\paragraph{Intervention.}
We add a one-sided, \emph{reference-anchored margin}. For a user $u$, let
$\mathcal{T}(u)$ be the set of valid logged targets and $\mathcal{G}(u)$ the
sampled candidates. Define a target's \emph{competitive gap} under a model $x$ as
its log-probability advantage over the best sampled candidate,
\begin{equation}
m_x(y;u) \;=\; \log P_x(y \mid u) \;-\; \max_{\hat{y}\in\mathcal{G}(u)} \log P_x(\hat{y}\mid u),
\qquad x \in \{\theta, \text{ref}\},
\label{eq:gap}
\end{equation}
and penalize the policy only when its gap on a logged target falls below the
reference model's gap:
\begin{equation}
\mathcal{L}_{\text{margin}}(\theta)
\;=\; \frac{1}{|\mathcal{T}(u)|}\sum_{y\in\mathcal{T}(u)}
      \big[\, m_{\text{ref}}(y;u) - m_{\theta}(y;u) \,\big]_{+},
\qquad [z]_+ = \max(z,0).
\label{eq:margin}
\end{equation}
The final objective is
\begin{equation}
\mathcal{L}_{\text{\mgrpo{}}}(\theta) \;=\; \mathcal{L}_{\text{GRPO}}(\theta) \;+\; \lambda\,\mathcal{L}_{\text{margin}}(\theta).
\label{eq:mgrpo}
\end{equation}
The margin is inactive wherever post-training has not eroded a logged target's
competitive position, so it never fights a reward improvement that leaves recall
intact; it activates precisely when the policy begins to trade a real target away
for a higher-reward sample. Because it is anchored to the reference model rather
than to a fixed constant, it scales naturally with how confidently the base model
already ranked each target. The encoder, embeddings, normalization state, and MHP
remain frozen throughout.

\paragraph{Result.}
Table~\ref{tab:mgrpo} is a one-factor $2\times 2$ on the same base checkpoint,
data, steps, and evaluation set. Neither term alone moves reward-recall: GRPO
without the margin is flat on reward-recall@10 ($-0.1\%$) while recall@10 decreases ($-1.25\%$), indicating a loss in recall without an
increase in reward-recall, and the margin alone
preserves recall but produces only limited reward movement ($+0.5\%$). Together
they improve both: $+1.6\%$ reward-recall@10 with recall@10 at $+0.3\%$ on one
node, and $+3.1\%$ / $+0.9\%$ on two nodes.
The PPO term optimizes the reward objective, while the margin penalizes decreases
in logged targets' competitive position. The GRPO-only and margin-only variants
are ablations of the joint objective, used to assess why both components are
needed for reward-weighted recall improvement with recall preservation. The
additional gain from training on more nodes comes from using more effective
training data.

\begin{table}[t]
\centering
\caption{Post-training ablation and scaling (one-factor $2\times 2$ on a common
base checkpoint and evaluation set; changes relative to the base model). Vanilla
GRPO performs worse than the base model on recall; \mgrpo{} improves reward-recall
while holding recall.}
\label{tab:mgrpo}
\small
\resizebox{\textwidth}{!}{%
\begin{tabular}{lccl}
\toprule
Training objective & Reward Recall@10 vs.\ base & Recall@10 vs.\ base & Readout \\
\midrule
GRPO only                    & $-0.1\%$ & $-1.25\%$ & Policy moved without reward gain \\
Margin only                  & $+0.5\%$ & $+0.4\%$  & Recall preserved; limited reward movement \\
GRPO + margin (\mgrpo{}), 1 node & $+1.6\%$ & $+0.3\%$ & Joint objective improves both \\
GRPO + margin (\mgrpo{}), 2 nodes & $+3.1\%$ & $+0.9\%$ & Measured Pareto point \\
\bottomrule
\end{tabular}}
\end{table}

\section{Deployment and Serving}
\label{sec:deploy}

\subsection{Serving architecture}
\label{sec:serving-arch}

\paragraph{Packaging and inference service.}
An inference-only artifact---the dense weights plus the embedding tables, with
optimizer and training-only state stripped---is packaged into a self-contained
model archive for online inference. The served model
runs only three components on the request path---the encoder over user context,
the block-wise decoder that generates Semantic IDs, and the MHP module that scores
them---so no dense item tower or maintained ANN index is needed online. Concurrent
requests are dynamically batched, and requests sharing generation settings are
grouped so each runs as one batched forward pass.

\paragraph{Per-request generation and scoring.}
A request carries only user-side features (no item features): the handler encodes
the user context, runs block-wise beam search to generate the top-$K$ Semantic-ID
candidates (with optional diversity control across the coarse hierarchy level), and
scores each candidate with the MHP module in the same forward pass. It returns the
candidate Semantic IDs together with their per-head engagement scores---generation
and ranking in a single service call.

\paragraph{Semantic-ID to item reverse mapping.}
A generated Semantic ID is not itself servable content; it must be mapped back to
concrete items. This mapping is maintained \textbf{asynchronously}, off the request
hot path: as new content is ingested it is tokenized and added to a catalog of
SID$\rightarrow$item assignments that is updated so that freshly generated SIDs resolve to current items
(Section~\ref{sec:freshness}). A single Semantic ID fans out to multiple items, so
the candidate pool for a request is the union of items over the generated SIDs,
after quality and eligibility filtering.

\paragraph{Ranking-stage control.}
Because generation is conditioned on user context alone and directly emits scored
candidates, a single service collapses the retrieval and early-ranking stages of a
classic funnel. \grp{} is deployed behind a configurable funnel that can
\textbf{bypass those stages} entirely or replace only a subset, injecting the
generative candidates at a chosen point of the existing stack, and that can let a
configurable portion of \grp{} candidates bypass the early-ranking stage. This is the
control surface for Phases~2 and~3 of Section~\ref{sec:progressive}: it isolates
the effect of generative retrieval from downstream ranking, and it lets a
production ranker stay in the loop while the unified model matures.

\subsection{Reducing end-to-end latency}
\label{sec:latency}

A generative retriever is on the request's critical path, so its latency
determines whether it can hold a quota at all. Instrumenting the full path
exposed bottlenecks on both sides of the service boundary.

\paragraph{Feed-orchestration side.}
(i) \emph{Event-sequence cap}: the orchestrator forwarded a user's entire event
history although the model consumes only a fixed event budget; capping the payload at the
model's budget reduced p99 request-body size by approximately 67--75\%.
(ii) \emph{Event flattening}: converting the typed event history into the model's
request format dominated orchestrator time because of an inefficient
implementation; a rewrite reduced it by approximately 96\%.
(iii) \emph{Reverse lookup}: the per-SID point reads that map generated Semantic
IDs to items were replaced by a single batched read against the key-value store,
cutting lookup p95 by approximately 70\% before caching. Together these
reduced the orchestrator-side p95 contribution by approximately 75\% or more.

\paragraph{Model-server side.}
(i) \emph{Preprocessing}: Python feature preprocessing dominated the model server;
rewriting it in C++ reduced preprocessing by approximately 37\%.
(ii) \emph{KV cache and CUDA graphs}: enabling a key--value cache during block-wise
decoding removes repeated computation across decode steps, and capturing the
decode graph once with CUDA graphs removes per-request graph construction;
together they cut generation-stage latency by approximately 85\%.
(iii) \emph{Worker tuning}: the model server's RPC queue limited throughput; two
workers with batch size 8 reduced RPC p95 by approximately 50\%.

\begin{table}[t]
\centering
\caption{Relative changes from inference optimizations in the evaluated
configuration. Component measurements and end-to-end latency use different
measurement boundaries; their changes are not additive.}
\label{tab:latency}
\small
\begin{tabular}{lc}
\toprule
Measurement & Relative change \\
\midrule
Model-server latency (p50), cumulative optimizations & $-67.7\%$ \\
Model-server throughput, cumulative optimizations & $+208\%$ \\
Generation-stage latency, KV cache and graph capture & $-84.6\%$ \\
Input-conversion latency & $-96.2\%$ \\
Model-server request latency (p95) & $\approx -50\%$ \\
Batched item-lookup latency (p95), before caching & $\approx -70\%$ \\
\midrule
End-to-end retrieval-stage latency (p95) & $-69.0\%$ \\
\bottomrule
\end{tabular}
\end{table}

\section{Online A/B Test}
\label{sec:ab}

The progressive path of Section~\ref{sec:progressive} gives three axes along
which the end-to-end model can take over the funnel: it can be a better
\emph{retrieval source}, its candidates can \emph{bypass the rankers} on its own
scores, and it can \emph{absorb the quota} of sources it beats. The axes are
independent knobs, so we explored them in parallel rather than in sequence, and
each has its own success criterion. As a retrieval source, the model is evaluated on the reported surface metrics;
the comparison tests how reward-guided post-training changes those outcomes. For the ranking bypass the goal is a
larger share of served traffic for \grp{} at neutral topline metrics: the bypass
removes a production ranker from part of the slate, so holding the topline is
the bar. For bypass combined with source replacement the goal is both at once:
a larger share \emph{and} a topline gain.

\subsection{Setup}
\label{sec:ab-setup}

All experiments run on the short-video surface and cover short-video content
only. Every treatment adds \grp{} as an online retrieval source, served live
through the stack of Section~\ref{sec:deploy}; the model variants are supervised
pre-training (SFT), reward-guided post-training (RL) using the view-time prediction head in the
reported comparison, and a larger per-request decoding budget. Results report relative
changes in the metrics of Section~\ref{sec:protocol}: the views, view
time, and shares on the surface, and \grp{}'s \emph{source-rate rank}, its rank
among the evaluated retrieval sources by share of served views. Significance is assessed at the
95\% level and bold marks $p<0.05$ in the tables. Topline guardrails, such as
DAUs and app-level engagement, remain neutral or positive in every configuration.

\subsection{Axis 1: \grp{} as a retrieval source}
\label{sec:ab-m1}

Here \grp{} sits behind the unchanged rankers with a fixed quota, and the
per-source protocol benchmarks it against the incumbent sources while we improve
the model. The supervised model enters as the 14th-ranked source by served share
with neutral surface-level metrics (Table~\ref{tab:ab-axis1}): views $0.00\%$, view
time $+0.04\%$, shares $+0.58\%$, none significant. This is the expected
starting point for a new source behind an intact funnel, and it establishes the
SFT model as the baseline against which model improvements are read.

Reward-guided post-training is the first such improvement. Measured against the
SFT model with the serving configuration held fixed, RL lifts \grp{}'s view-time
contribution by $+0.45\%$; view time is the sole reward used for post-training
(Section~\ref{sec:posttrain}). Adding a larger
per-request decoding budget, made affordable by the serving work of
Section~\ref{sec:deploy}, raises the source to 5th by served share and turns the
surface-level results positive against production: $+0.39\%$ in view time
($p=.031$) and $+0.31\%$ in views ($p=.091$), with only the view-time gain
significant. Raising the budget further to $20\times$ moves the source to 4th by
served share and makes both view time ($+0.46\%$) and shares ($+0.77\%$)
significant, with views at $+0.22\%$, indicating that the retrieval-only
configuration continues to benefit from a larger decoding budget.

\begin{table}[t]
\centering
\caption{Axis 1, \grp{} as a retrieval source behind the unchanged rankers.
Relative changes on the short-video surface against the stated baseline; bold
marks $p<0.05$. The RL row is measured against the SFT model with the same
serving configuration; its views and shares were not part of that read-out.}
\label{tab:ab-axis1}
\small
\setlength{\tabcolsep}{5pt}
\begin{tabular}{llcccc}
\toprule
Model & Baseline & Source-rate rank & Views & View time & Shares \\
\midrule
SFT                       & production & 14th & $0.00\%$  & $+0.04\%$          & $+0.58\%$ \\
SFT + RL                  & SFT        & ---  & ---       & $+0.45\%$          & --- \\
SFT + RL + $4\times$ budget  & production & 5th  & $+0.31\%$ & $\mathbf{+0.39\%}$ & $+0.29\%$ \\
SFT + RL + $20\times$ budget & production & 4th & $+0.22\%$ & $\mathbf{+0.46\%}$ & $\mathbf{+0.77\%}$ \\
\bottomrule
\end{tabular}
\end{table}

The per-source benchmark says where the model stands among the sources it
competes with. Table~\ref{tab:ab-source} compares the post-trained model's
outcome profile with the all-source average in the same treatment (measured in
the early-bypass configuration of Section~\ref{sec:ab-m2}, which is also the base
for source replacement). \grp{} ranks 2nd on completion and on sends, improves
average watch time and reduces skips relative to the average, and is
above average on favorites; its watch-time and skip ranks leave room against the
strongest individual sources. These comparisons identify the remaining differences between sources.

\begin{table}[t]
\centering
\caption{Per-source engagement for SFT + RL with early-ranking bypass,
the base configuration for source replacement, relative to the all-source average
in the same treatment. Parentheses give ranks among sources with at least 1\%
of the benchmark's attributed observations and a defined metric.
The cutoff applies only to ranks.
Completion is the mean watch-time-to-video-length ratio over valid observations;
favorite and send rates measure the fraction of attributed observations with at
least one such event. Lower skip rate is better. These estimates attribute engagement to served
recommendations and can overcount viewed items.}
\label{tab:ab-source}
\small
\setlength{\tabcolsep}{10pt}
\begin{tabular}{lc}
\toprule
Metric & Relative improvement (rank) \\
\midrule
Average watch time & $+4.9\%$ (12th) \\
Completion ratio   & $+14.6\%$ (2nd) \\
Skip rate (lower is better) & $-10.1\%$ (10th) \\
Favorite rate      & $+9.2\%$ (9th) \\
Send rate          & $+34.9\%$ (2nd) \\
\bottomrule
\end{tabular}
\end{table}

\subsection{Axis 2: bypassing early ranking}
\label{sec:ab-m2}

The bypass lets a configurable portion of \grp{}'s candidates skip the
early-ranking stage and travel on the model's own MHP scores
(Section~\ref{sec:serving-arch}). Because a production ranker is removed from
part of the slate, the goal is to \emph{raise \grp{}'s share of served traffic
while holding the topline neutral}; a topline gain is desirable, although the acceptance criterion is neutrality.
Table~\ref{tab:ab-bypass} reports the three configurations, each against
production.

\begin{table}[t]
\centering
\caption{Axis 2, early-ranking bypass, against production. Source-rate rank is
\grp{}'s rank among the evaluated sources by share of served views (SFT without
bypass ranks 14th); the remaining columns are relative changes on the short-video
surface, bold marking $p<0.05$.}
\label{tab:ab-bypass}
\small
\setlength{\tabcolsep}{5pt}
\begin{tabular}{lcccc}
\toprule
Configuration & Source-rate rank & Views & View time & Shares \\
\midrule
SFT + bypass                       & 8th & $-0.16\%$ & $\mathbf{-0.45\%}$ & $\mathbf{+2.87\%}$ \\
SFT + RL + bypass                  & 9th & $+0.22\%$ & $+0.04\%$          & $\mathbf{+0.98\%}$ \\
SFT + RL + bypass + larger budget  & 4th & $+0.24\%$ & $-0.24\%$          & $\mathbf{+2.30\%}$ \\
\bottomrule
\end{tabular}
\end{table}

With bypass, the supervised model moves from 14th to 8th by served-share rank,
and with post-training and a larger decoding budget \grp{} reaches 4th. Whether the surface-level metrics stay neutral depends on
the model. With the supervised model, shares rise $+2.87\%$ while view time falls
$-0.45\%$, both significant. These results show different effects on sharing and view time. The post-training
comparison evaluates whether reward alignment changes that trade-off before the
generative model is used for ranking. Post-training the same model with \mgrpo{} against the frozen
in-model reward closes the gap: view time is flat ($+0.04\%$) and shares stay
positive ($+0.98\%$), which meets the neutrality bar at a higher share than SFT
without bypass. The larger decoding budget then yields a served-share rank of 4th while the
surface-level results remain neutral (view time $-0.24\%$, $p=.177$; shares $+2.30\%$).
Together with Table~\ref{tab:ab-axis1}, these results distinguish the deployment
effects: bypass increases served share and sharing, with a view-time cost that
post-training offsets, while a larger decode budget improves view time in the
retrieval-only configuration. Extending the bypass toward the
late-stage ranker requires the in-model ranking quality of
Section~\ref{sec:reward-fix} to be demonstrated online.

\subsection{Axis 3: bypass with retirement of weaker sources}
\label{sec:ab-m3}

The third axis combines the early-ranking bypass with the retirement of a few
low-performing incumbent sources, whose quota is reassigned to \grp{}. The goal is both a larger share of served traffic and a gain in the surface-level metrics. Applied to
the post-trained early-bypass configuration, the treatment improves the surface-level
metrics (Table~\ref{tab:ab-replace}): $+0.47\%$ views, $+0.82\%$
view time, and $+2.56\%$ shares against production, all significant, at neutral
guardrails, where the base configuration alone was neutral on views and view
time. \grp{} ranks 10th by served share in this experiment, so the measured gain
came with the served share holding rather than growing; the served-share rank is
measured within each experiment's own arm and is not directly comparable across
rows. The gain is accompanied by an expected shift in the
content mix---the surface absorbs slots previously served from other content
types---without adverse changes in the reported surface metrics, with neutral
platform-level guardrails.

\begin{table}[t]
\centering
\caption{Axis 3, early-ranking bypass combined with retirement of low-performing
sources, against production. The first row is the base configuration (Axis 2,
Table~\ref{tab:ab-bypass}); the second adds the source retirement. Source-rate
rank is \grp{}'s rank among the evaluated sources by share of served views; the
remaining columns are relative changes on the short-video surface, bold marking
$p<0.05$.}
\label{tab:ab-replace}
\small
\setlength{\tabcolsep}{5pt}
\begin{tabular}{lcccc}
\toprule
Configuration & Source-rate rank & Views & View time & Shares \\
\midrule
SFT + RL + bypass                        & 9th  & $+0.22\%$          & $+0.04\%$          & $\mathbf{+0.98\%}$ \\
SFT + RL + bypass + source retirement    & 10th & $\mathbf{+0.47\%}$ & $\mathbf{+0.82\%}$ & $\mathbf{+2.56\%}$ \\
\bottomrule
\end{tabular}
\end{table}

\subsection{Where the path stands}
\label{sec:ab-status}

The three axes evaluate complementary parts of the progressive deployment
strategy. As a source, \grp{} is competitive on a few metrics
(2nd on completion and sends), and post-training improves view time, the metric
used as its reward. The early-ranking bypass raises its share of served traffic from 14th to 4th
by rank, and post-training turns the bypass from a view-time loss into neutral
surface-level results. Retiring weaker sources on top of the bypass converts the share into
significant gains in views, view time, and shares at neutral guardrails. The
remaining gaps are the ones the benchmark names: average watch time and skip
rate still trail the strongest individual sources, and the late-stage ranker has
not yet been bypassed. Further evaluation of quota changes and ranking-stage bypass should consider the
reported outcome trade-offs together with the model's ranking quality.

\section{Conclusion, Limitations, and Future Directions}
\label{sec:conclusion}

This report presented \grp{}, a generative recommendation paradigm that unifies
retrieval, ranking, and reward modeling in a single model, and the progressive
path by which we are bringing it into production. Retrieval is generative decoding
of Qwen3-VL Semantic IDs from a decoder-heavy encoder--decoder trunk with
block-wise-independent target decoding and variable-length attention over
event histories; ranking is a jointly-trained, detached MHP module that
attaches calibrated engagement scores to candidates; and reward modeling reuses
that same frozen module for \mgrpo{} post-training. On the pre-training side we
showed that item-level history fusion, decoder-heavy capacity, a merged stream
of explicit actions with reserved budget, and a higher-resolution tokenizer each
improve retrieval
in one-factor comparisons. On the post-training side we diagnosed why a naively
trained reward is candidate-indifferent, fixed it, and introduced a
reference-anchored margin that lets reward optimization proceed without eroding
recall. On the serving side we reduced end-to-end retrieval latency by 69\%.
Online, the post-trained early-bypass configuration delivers strong completion
and sharing alongside above-average watch time, RL steers \grp{} toward the
metric it is rewarded on, and
replacing weaker sources with it yields significant surface-level gains at neutral
guardrails.

The distance from this system to a fully end-to-end recommender is also clear,
and the progressive protocol measures it directly.

\begin{enumerate}[leftmargin=1.8em]
\item \textbf{Not yet leading on every axis.} Despite stronger watch time,
  completion, and skip performance than the all-source average, the post-trained
  early-bypass configuration remains below average on boost rate and trails the
  strongest individual sources on watch time and skip rate. Auxiliary sources
  remain necessary, and quota growth is bounded by these gaps rather than by
  infrastructure. \emph{Next:} multi-objective rewards that include these axes,
  so that RL optimizes engagement along multiple dimensions rather than view time
  alone.
\item \textbf{Rankers are still in the loop, and the in-model ranker is not yet
  good enough to replace them.} Only a portion of \grp{} candidates bypasses the
  early-ranking stage and every candidate still passes the late-stage ranker;
  with the supervised model the bypass trades view time for sharing, and
  post-training only brings it back to neutral. The MHP module is the binding
  constraint: even after the redesign of Section~\ref{sec:reward-fix} it trails
  the production ranker, and the gap is widest on the rare heads (favorite,
  share, reply), where AUPRC is low in absolute terms and the
  reward is therefore least informative about the actions that matter most.
  \emph{Next:} close the MHP gap on the served distribution---more capacity and
  training data for the rare heads, and calibration monitoring against the
  production ranker.
\item \textbf{The model is small.} The deployed trunk is a relatively small dense
  model, and its size shows: histories beyond $B_{\mathrm{ref}}$ events are flat to negative,
  RL gains grow with batch scale, and the encoder--decoder rebalancing of
  Section~\ref{sec:encdec} improved quality and throughput at the same time,
  which suggests the dense capacity is under-provisioned rather than misallocated.
  \emph{Next:} evaluate larger dense models, verify that
  the generative-recommendation loss follows a predictable scaling law as reported
  for other production systems, and re-run the history-length and post-training
  ladders at each size to confirm that capacity is what unlocks longer histories
  and larger RL gains.
\end{enumerate}

We believe the central lesson generalizes beyond our platform: end-to-end
generative recommendation becomes production-viable not by replacing the cascade
with a bare generator, but by absorbing ranking and reward modeling \emph{into}
the generative model, by respecting the non-stationarity of the catalog with
continuous retraining, and by migrating progressively, one measurable source and
stage at a time.

\bibliographystyle{abbrvnat}
\bibliography{references}

\clearpage
\appendix
\section*{Appendix}
\addcontentsline{toc}{section}{Appendix}

\section{Contributions}
\label{app:contrib}

\textbf{Contributors}

\vspace{0.4em}
\noindent
\begin{minipage}[t]{0.32\textwidth}
Wenfeng Zhuo\\
Vincent Xue\\
Charles Wei\\
Cong Ni\\
Ruiming Lu\\
Jiwen Ren\\
Mo Li\\
Peng Yang
\end{minipage}%
\begin{minipage}[t]{0.32\textwidth}
Xufei Wang\\
Dongheng Li\\
Jiacong He\\
Yi Song\\
Yufei Fan\\
Mikhail Obukhov\\
Yiwen Chen
\end{minipage}%
\begin{minipage}[t]{0.32\textwidth}
Yvette Liu\\
Yin Ye\\
Chengjie Wu\\
Mingtao Zhang\\
Jinchao Ye\\
Lili Zhang\\
Chunhui Zhu
\end{minipage}%

\paragraph{Use of AI tools.}
AI tools assisted with manuscript editing. All methodologies and supporting
evidence were developed by humans and reviewed by the authors, who take
responsibility for the content of this report.

\section{Extended Related Work}
\label{app:related}

Generative recommendation can be framed as a sequence-modeling problem: a user's
interaction history is consumed, and the model \emph{generates} the identifier of
the next item token-by-token rather than scoring items from a fixed corpus.
Different systems realize this differently---encoder--decoder or decoder-only
backbones, item-level or Semantic-ID targets, retrieval-only or
retrieval-plus-ranking scope. We organize prior work into three strands and
position \grp{} by how far it unifies all three.

\paragraph{Generative retrieval with Semantic IDs.}
TIGER~\citep{tiger2023} introduced \emph{Semantic IDs}---item identifiers obtained
by residual-quantizing (RQ-VAE) a content embedding---and framed retrieval as
sequence-to-sequence generation of these IDs, showing a single model can retrieve
from a corpus without a maintained ANN index. PLUM~\citep{plum2026} adapts
pre-trained language models to the same generative-retrieval formulation at
industrial scale via Semantic-ID tokenization and continued pre-training. This
line establishes the tokenization and decoding machinery \grp{} builds on, but
targets retrieval alone.

\paragraph{End-to-end generative recommendation.}
OneRec~\citep{onerec2024,onerec_tr2025} is the closest precedent: it unifies
retrieval and ranking in one encoder--decoder generative recommender with
\emph{iterative preference alignment}, and OneRec-V2~\citep{onerecv2_2025}
moves to a lazy decoder-only architecture and reinforcement learning against real
online feedback. UniPinRec~\citep{unipinrec2026} pushes toward a full-stack
unification of retrieval and ranking at industrial scale within an existing serving
stack, and GPR~\citep{gpr2025} recasts large-scale advertising recommendation as a
single generative ``one-model'' paradigm. These systems validate that a generative
recommender can be aligned with preference signals; \grp{} differs in that the
reward used for alignment is \emph{the model's own jointly-trained ranking module},
frozen and reused, rather than an externally or separately trained preference
model, and in the explicitly progressive deployment path by which it displaces the
incumbent cascade.

\paragraph{Generative modeling for ranking.}
HSTU~\citep{hstu2024} recasts ranking and retrieval as a single \emph{generative
transducer} over a unified action sequence and demonstrates trillion-parameter
scaling with hardware-efficient attention, but keeps retrieval and ranking as
distinct downstream heads. Gryphon~\citep{gryphon2026} is closest in spirit to
\grp{}'s ranking design: it augments an encoder--decoder Semantic-ID generator
with a jointly-trained item-level scoring head computed in a single forward pass.
\grp{} extends this design by \emph{reusing} that same jointly-trained head,
frozen, as the reinforcement-learning reward that aligns the generator.

\paragraph{Policy optimization.}
\mgrpo{} builds on GRPO~\citep{grpo2024}, itself a critic-free variant of
PPO~\citep{ppo2017}. Its margin term is related in spirit to preference-margin
objectives, but is anchored to the reference model's competitive gap on
\emph{logged targets} rather than to pairwise preferences over sampled
candidates, which is what allows it to protect recall specifically. We also draw
on standard components: DCNv2~\citep{dcnv2_2021} feature crossing, rotary position
embeddings~\citep{rope2021}, FlashAttention~\citep{flashattention2022}, the
Muon~\citep{muon2024} and Lamb~\citep{lamb2020} optimizers, and
Qwen3-VL~\citep{qwen3vl2025} embeddings for tokenization.

\end{document}